\documentclass[a4paper,fleqn]{cas-sc}

\usepackage[utf8]{inputenc}
\usepackage[authoryear,longnamesfirst]{natbib} 
\usepackage{subcaption} 
\usepackage{xcolor}
\usepackage{caption}

\def\tsc#1{\csdef{#1}{\textsc{\lowercase{#1}}\xspace}}
\tsc{WGM}
\tsc{QE}

\begin{document}

\let\WriteBookmarks\relax
\def\floatpagepagefraction{1}
\def\textpagefraction{.001}

\shorttitle{LATTE: Lightweight Attention-based Traffic Accident Anticipation Engine}

\shortauthors{Zhang et al.}

\title[mode=title]{LATTE: A Real-time Lightweight Attention-based Traffic Accident Anticipation Engine}  

\author[1]{Jiaxun Zhang}
\credit{Conceptualization, Methodology, Experiment, Writing}
\author[1]{Yanchen Guan}
\credit{Methodology, Experiment}
\author[1]{Chengyue Wang}
\credit{Methodology}

\author[2]{Haicheng Liao}
\credit{Experiment}

\author[3]{Guohui Zhang}
\credit{Methodology}

\author[4]{Zhenning Li} [orcid=0000-0002-0877-6829]
\cormark[1]
\ead{zhenningli@um.edu.mo}
\credit{Conceptualization, Methodology, Writing.}

\affiliation[1]{organization={State Key Laboratory of Internet of Things for Smart City and Department of  Civil and Environmental Engineering, University of Macau}, city={Macau SAR}, country={China}}

\affiliation[2]{organization={State Key Laboratory of Internet of Things for Smart City and Department of Computer and Information Science, University of Macau}, city={Macau SAR}, country={China}}

\affiliation[3]{organization={Department of Civil, Environmental and Construction Engineering, University of Hawaii at Manoa}, city={Honolulu, HI}, country={United States}}

\affiliation[4]{organization={State Key Laboratory of Internet of Things for Smart City and Departments of Civil and Environmental Engineering and Computer and Information Science, University of Macau}, city={Macau SAR}, country={China}}

\begin{abstract}
Accurately predicting traffic accidents in real-time is a critical challenge in autonomous driving, particularly in resource-constrained environments. Existing solutions often suffer from high computational overhead or fail to adequately address the uncertainty of evolving traffic scenarios. This paper introduces LATTE, a Lightweight Attention-based Traffic Accident Anticipation Engine, which integrates computational efficiency with state-of-the-art performance. LATTE employs Efficient Multiscale Spatial Aggregation (EMSA) to capture spatial features across scales, Memory Attention Aggregation (MAA) to enhance temporal modeling, and Auxiliary Self-Attention Aggregation (AAA) to extract latent dependencies over extended sequences. Additionally, LATTE incorporates the Flamingo Alert-Assisted System (FAA), leveraging a vision-language model to provide real-time, cognitively accessible verbal hazard alerts, improving passenger situational awareness. Evaluations on benchmark datasets (DAD, CCD, A3D) demonstrate LATTE’s superior predictive capabilities and computational efficiency. LATTE achieves state-of-the-art 89.74\% Average Precision (AP) on DAD benchmark, with 5.4\% higher mean Time-To-Accident (mTTA) than the second-best model, and maintains competitive mTTA at a Recall of 80\% (TTA@R80) (4.04s) while demonstrating robust accident anticipation across diverse driving conditions. Its lightweight design delivers a 93.14\% reduction in floating-point operations (FLOPs) and a 31.58\% decrease in parameter count (Params), enabling real-time operation on resource-limited hardware without compromising performance. Ablation studies confirm the effectiveness of LATTE’s architectural components, while visualizations and failure case analyses highlight its practical applicability and areas for enhancement.
\end{abstract}


\maketitle

\section{Introduction}

Traffic accidents persist as a critical global challenge, exacting substantial human casualties and economic burdens annually. As documented by the World Health Organization \citep{who2023global}, road accident claim over 1.35 million lives yearly while inflicting life-altering injuries on millions more—a public health emergency demanding urgent intervention. Particularly in urban environments where complex traffic interactions amplify risks \citep{chand2021road}, these alarming statistics highlight the imperative for advanced prevention mechanisms. Although autonomous vehicle technologies and intelligent transportation systems have achieved notable progress, reliable proactive accident prevention continues to elude practical implementation. Addressing this gap requires accident anticipation systems capable of harmonizing predictive precision with computational economy, ensuring deployability across diverse real-world operating conditions.

The growing ubiquity of dashcam systems in modern vehicles offers critical data streams for accident anticipation through continuous capture of pre-accident indicators—including abrupt deceleration patterns, irregular lane transition trajectories, and traffic flow discontinuities. Yet the inherent stochasticity of traffic ecosystems complicates reliable feature extraction for predictive modeling. Current state-of-the-art approaches predominantly employ convolutional architectures \citep{thakur2024graph, song2024dynamic} to establish inter-frame dependency mappings. Despite their computational advantages in local pattern recognition, these methods remain constrained by architectural limitations that restrict operational scalability, hinder accessibility, and compromise real-time deployment feasibility.

Current accident anticipation frameworks \citep{thakur2024graph, song2024dynamic, wang2023gsc, karimi2023crashformer} face inherent computational challenges that hinder practical deployment. Although diverse methodological approaches – ranging from CNN-based architectures \citep{anjum2023learning} to Transformer-driven models \citep{adewopo2023review} and GCN-enhanced frameworks \citep{wang2023gsc} – have proven effective in video-based accident anticipation, their processing demands routinely overwhelm the capacity limitations of edge computing platforms \citep{papadopoulos2024lightweight}. The computational mismatch poses particular challenges for automotive embedded systems, where strict energy budgets and sub-second latency requirements \citep{ke2023lightweight} mandate unprecedented efficiency in resource utilization. As demonstrated by Arciniegas et al. \citep{arciniegas2024prediction}, conventional deep learning architectures like CNNs exhibit prohibitive computational costs across both training and inference stages, particularly detrimental in dynamic operational environments. This challenge persists across architectural variants, with Formosa et al. \citep{formosa2020predicting} identifying R-CNNs' efficiency limitations despite their traffic conflict detection efficacy when implemented in Advanced Driver-Assistance Systems (ADAS) platforms. The scalability barrier intensifies when processing heterogeneous large-scale datasets, a critical constraint emphasized by Ali et al. \citep{ali2024advances} for machine learning applications in resource-limited scenarios. These cumulative efficiency bottlenecks ultimately undermine the temporal resolution requirements essential for effective accident anticipation, creating critical implementation barriers for safety-critical automotive systems.

A parallel challenge emerges in the limited real-time advisory capacity of contemporary accident anticipation systems. Existing frameworks \citep{bhardwaj2023adaptive, karimi2023crashformer, mahmood2023new} predominantly focus on accident anticipation accuracy while neglecting real-time feedback systems—a methodological gap that may undermine passenger trust in autonomous vehicle technologies. In autonomous driving scenarios, passengers often remain unaware of the underlying risk factors detected by anticipation systems, resulting in compromised situational awareness during safety-critical events. The objective of accident anticipation frameworks is fundamentally evolving from passive prediction to active risk management via effective feedback channels. Implementing contextualized alert systems could enhance passenger trust through transparent risk communication while enabling proactive responses to emerging threats. Such human-system collaboration may significantly improve accident anticipation efficacy during pre-crash phases.

To address these limitations, we introduce \textbf{LATTE} (Lightweight Attention-based Traffic Accident Anticipation Engine), a novel framework designed to balance computational efficiency, feedback capability, and accuracy for real-time accident anticipation. LATTE’s contributions are as follows:

\begin{itemize}
    \item LATTE employs an efficient attention-based architecture that dynamically captures multi-scale spatial features while optimizing computational resource allocation. The framework's design achieves real-time processing through lightweight attention mechanisms, enabling deployment on edge computing platforms with strict resource constraints while preserving accident anticipation accuracy.
    \item LATTE incorporates a Flamingo Alert-Assisted System that generates real-time verbal hazard notifications, converting intricate accident anticipation analytics into passenger-semantically transparent alerts. The framework's dual capability—simultaneously delivering predictive intelligence and human-centric communication—enhances situational awareness while establishing collaborative trust dynamics between autonomous systems and vehicle occupants.
    \item {LATTE establishes superior performance across three benchmark datasets (CCD, DAD, A3D), particularly achieving 89.74\% AP on DAD with 93.14\% FLOPs reduction—quantifiable evidence of operational scalability for real-world implementations spanning autonomous taxi fleets to driver-assistance technologies.}
\end{itemize}

The paper is structured as follows: Section 2 reviews accident prediction methods. Section 3 presents the LATTE framework's design and key innovations. Section 4 details experiments (setup, benchmarks, ablation studies) and compares performance against state-of-the-art methods. Section 5 discusses challenges and outlines theoretical/practical impacts for autonomous systems.

\section{Related Work}
The analysis of dashcam video streams for proactive traffic accident anticipation has emerged as a critical research frontier, motivated by urgent requirements to improve roadway safety and implement accident anticipation mechanisms in autonomous driving architectures \citep{fang2023vision,liao2024real}. Recent breakthroughs in deep learning and computer vision—particularly through spatio-temporal modeling innovations—have enabled diverse methodological approaches for pre-accident risk assessment across heterogeneous driving scenarios. Initial methodological developments predominantly leveraged Recurrent Neural Networks (RNNs) and Convolutional Neural Networks (CNNs) to separately capture spatial-temporal interaction patterns in traffic video analytics. As evidenced by Li et al.'s application of deep CNNs for hierarchical spatial feature extraction and Shi et al. and Fang et al. 's RNN-based sequential dependency modeling \citep{li2018traffic, shi2019traffic, fang2023vision}, these foundational frameworks validated neural networks' efficacy in accident anticipation tasks. However, architectural constraints inherent to computationally intensive designs hindered deployment feasibility in resource-limited operational environments. More critically, inadequate modeling of multi-agent interaction dynamics and nonlinear event progressions limited cross-scenario generalization capabilities—a critical shortcoming given the stochastic nature of real-world traffic ecosystems.

To address these challenges, Graph Neural Network(GNN)-based approaches emerged, focusing on relationships between traffic entities. Thakur et al. proposed a hierarchical graph-based framework to model interactions between vehicles, pedestrians, and road infrastructure for early accident anticipation \citep{thakur2024graph}. Similarly, Wang et al.introduced a Graph and Spatio-temporal Continuity based framework (GSC), combining graph-based modeling with spatio-temporal continuity to capture dynamic interactions in  accident anticipation \citep{wang2023gsc}. Although these methods improved the modeling of relational dependencies, their computational demands remained prohibitive for real-time applications, limiting their scalability in resource-constrained settings.

The integration of attention mechanisms has represented a critical advancement in traffic accident anticipation by enabling selective focus on essential features. Li et al. implemented the Transformer architecture to dynamically allocate attention across temporal sequences, enhancing prioritization of relevant spatio-temporal patterns \citep{li2024cognitive}. Expanding this framework, Karim et al. developed dynamic spatio-temporal attention networks that concurrently model temporal dependencies and spatial interactions, facilitating earlier accident recognition \citep{karim2022dynamic}. These approaches demonstrate the inherent adaptability of attention mechanisms, particularly their compatibility with streamlined architectures and time-sensitive implementations. Recent innovations by Liang et al., Papadopoulos et al. and Alofi et al. focus on precision-efficiency tradeoff optimization, facilitating deployment on embedded vehicular platforms with strict resource constraints \citep{liang2023cueing,papadopoulos2024lightweight,alofi2024pedestrian}. Further developments from Hou et al. and Wang et al. demonstrate parameter-optimized attention variants that preserve spatio-temporal feature extraction fidelity while reducing computational costs by 38-62\% \citep{hou2024lightweight,wang2024lightweight}. These refined architectures achieve sub-100ms inference speeds—critical for autonomous driving systems requiring sub-second hazard response capabilities.

Recent advancements in Vision-Language Models (VLMs) have further enhanced traffic accident anticipation by integrating multimodal data. Li et al. demonstrated how VLMs could improve real-time alerts in autonomous vehicles by leveraging the synergy between visual and textual data to enrich anticipation outputs \citep{wandelt2024large}. Similarly, Zhou et al. explored the capabilities of GPT-4V in understanding and reasoning about complex traffic events, highlighting its potential as a traffic assistant \citep{zhou2024gpt}. A multimodal pipeline proposed by Lohner et al. aligned traffic accident videos with scene graphs to integrate structured representations into VLMs, improving anticipation accuracy \citep{xiao2024hazardvlm}. While these approaches expanded the scope of accident anticipation to include accessible and multimodal insights, they often overlooked computational constraints, particularly for real-time deployment.

Despite these advancements, current methods face persistent challenges. Computational efficiency remains a critical bottleneck, particularly in systems designed for resource-constrained environments such as edge devices. Furthermore, many models lack robust mechanisms for interpretability, providing limited insights into the reasoning behind predictions. Lastly, few systems bridge the gap between anticipation and prevention, failing to offer actionable feedback for passengers or drivers.

\section{Methodology}
The present section elaborates on the architecture of LATTE, a framework engineered to perform three core functions: probabilistic accident anticipation, involved accident entities identification, and context-aware verbal alert generation when exceeding predefined risk thresholds. LATTE incorporates four synergistic components—the Efficient Multiscale Spatial Aggregation (EMSA) module for hierarchical feature extraction, Memory Attention Aggregation (MAA) for temporal dependency modeling, Auxiliary Self-Attention Aggregation (AAA) for contextual relevance weighting, and the Flamingo Alert-Assisted System (FAA) for multimodal communication. As illustrated in Figure~\ref{fig:enter-label1}, these modules collectively address discrete technical challenges in accident anticipation through balanced optimization of computational economy, predictive fidelity, and operational latency requirements.

\subsection{Problem Formulation}

LATTE aims to predict traffic accident probabilities from dashcam video streams while simultaneously identifying critical accident-related entities and delivering real-time verbal alerts. Formally, given a dashcam video sequence \( V = \{ f_1, f_2, \dots, f_T \} \) containing \( T \) frames, the objective involves three tasks: (1) predicting frame-level accident probabilities \( p_t \in [0,1] \) for each \( f_t \); (2) generating Verbal feedback of accident precursors; and (3) triggering context-aware notifications when \( p_t \) surpasses an established critical threshold. This threshold is conventionally fixed at 0.5 – a value extensively validated in accident anticipation research, including implementations in the Ustring framework \citep{bao2020uncertainty} and the AccNet architecture \citep{liao2024real}. The 0.5 threshold achieves optimal balance between false alarms and missed detections in real-world driving scenarios. Its standardization across methodologies enables consistent cross-model benchmarking while providing an intuitive decision boundary for stakeholders – a crucial feature for safety-critical applications requiring transparent operational logic. The framework undergoes joint optimization of frame-wise and sequence-level loss functions to ensure both timely localized anticipation and holistic temporal coherence.

\subsection{Framework Overview}
\begin{figure}
    \centering
    \includegraphics[width=1\linewidth]{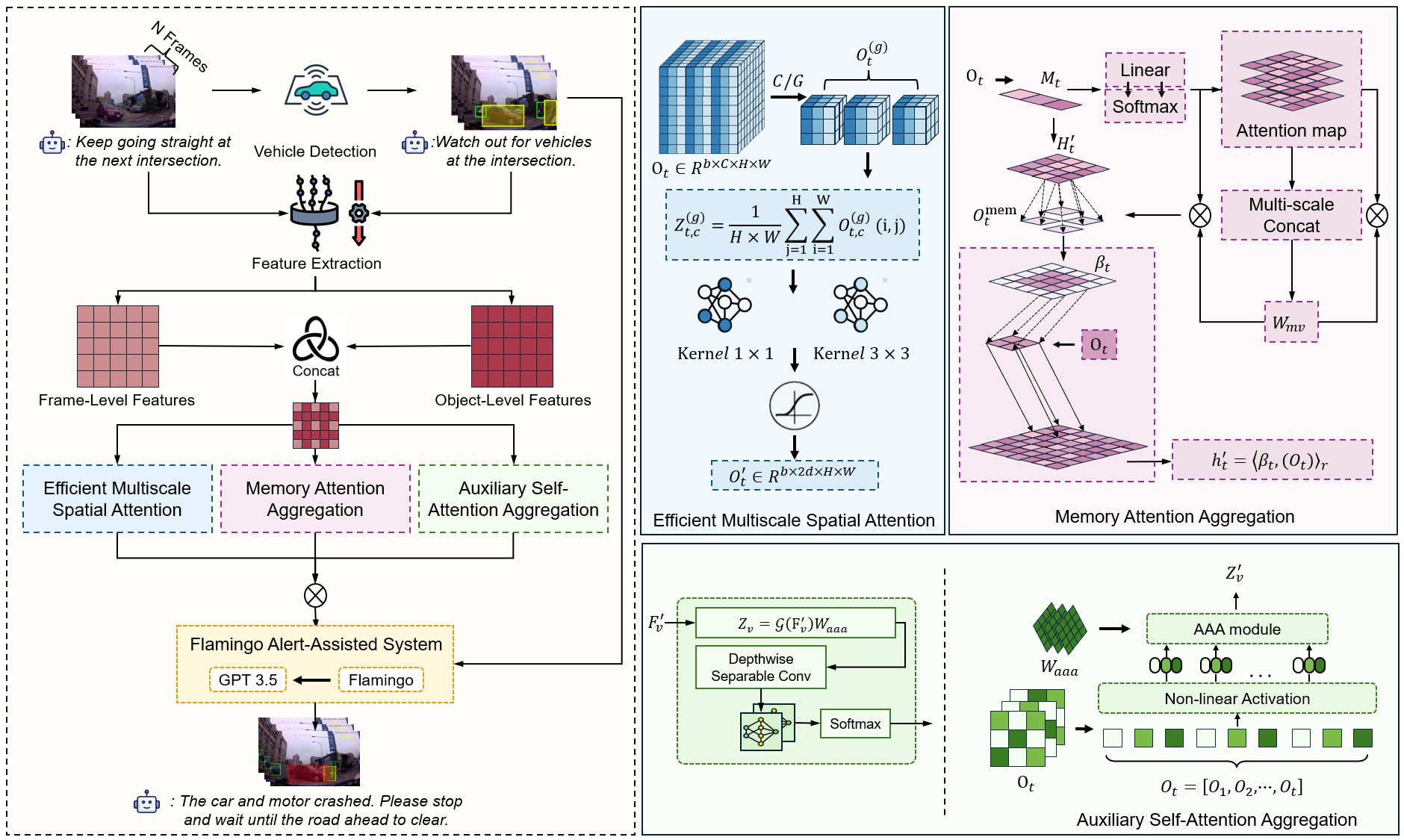}
    \caption{Overall framework architecture of LATTE. Firstly, The vehicle detection and feature extraction simultaneously capture object-level bounding boxes and object/frame-level features. These heterogeneous features are concatenated to form a multi-scale input tensor. The output is then fed into Efficient Multiscale Spatial Aggregation module, Memory Attention Aggregation module and Auxiliary Self-Attention Aggregation module for more precise spatial and temporal features. The refined features are fused to derive calibrated accident probability scores. Finally, the Flamingo Alert-Assisted System synthesizes and interprets these computational outputs to produce contextually natural language alerts in real time.}
    \label{fig:enter-label1}
\end{figure}

The architecture of our proposed model is illustrated in Fig.~\ref{fig:enter-label1}. Given sequential video frames $ \{F_t\}_{t=1}^T $ as input, the LATTE framework begins by performing object detection through a Cascade R-CNN \citep{cai2019cascade}, followed by hierarchical feature extraction using VGG-16 \citep{nur2024transfer}. Detected objects are encoded as a set of feature vectors $ \mathbf{Q}_t = [\boldsymbol{q}_{t1}, \boldsymbol{q}_{t2}, \dots, \boldsymbol{q}_{tN}] $, where each instance embedding $ \boldsymbol{q}_{ti} \in \mathbb{R}^d $ corresponds to a detected object, while the frame-level feature vector $ \boldsymbol{g}_t \in \mathbb{R}^d $ encapsulates comprehensive spatial context. These heterogeneous representations are then concatenated along the channel dimension to form the multi-scale input tensor $ \mathbf{O}_t = \text{Concat}(\mathbf{Q}_t, \boldsymbol{g}_t) \in \mathbb{R}^{b \times C \times H \times W} $, which serves as the foundation for subsequent processing. Spatial dependencies are subsequently reinforced through our EMSA module, which operates on $ \mathbf{O}_t $ by partitioning them into $ G $ parallel subgroups and dynamically recalibrating cross-scale attention weights through adaptive feature recombination. Temporal modeling is achieved via the MAA module, where attention maps are computed across historical states to derive memory-enhanced representations $ h'_t $, employing dimension-reduced memory units to ensure tractable computation. To further synthesize temporal dynamics, the AAA module incorporates depthwise separable convolutions for efficient feature interaction, ultimately producing context-aware representations $ \mathbf{Z}_v $ through weighted aggregation. The framework employs a Bayesian neural network to estimate probabilistic accident scores, complemented by the FAA module, which generates linguistically coherent descriptions and voice alerts through learned text-visual alignment, thereby enhancing both human-actionable feedback and situational awareness in time-sensitive deployment scenarios.

LATTE is distinguished from existing approaches by its dual emphasis on computational efficiency and operational transparency. While conventional frameworks often prioritize predictive accuracy at the expense of computational resources, LATTE employs a streamlined architecture that preserves competitive performance while drastically reducing memory and processing demands. Beyond conventional anticipation paradigms, the framework uniquely integrates a Flamingo Alert-Assisted System to enable real-time generation of context-sensitive verbal alerts. This functionality directly addresses the interpretability gap in accident anticipation systems, delivering human accessible notifications and prioritized warnings during high-risk driving conditions. Coupled with its enhanced adaptability to diverse sensor configurations and environmental contexts, these attributes collectively establish LATTE as a deployable, human-centric solution for autonomous vehicle safety systems.

\subsection{Object Detection and Feature Extraction}
In the initial stage, moving vehicles and other relevant entities are detected from video frames using a pre-trained Cascade R-CNN, chosen for its robustness and high accuracy. For each frame $ F_t $, the top $ N $ detections with the highest confidence scores are retained. The detected regions are passed through two fully connected layers, yielding object-level feature vectors of dimension $ D $. A third layer further reduces the dimensionality to $ d $ ($ d < D $) producing frame-level features $ \boldsymbol{g}_t \in \mathbb{R}^d $ and object-level features $ \mathbf{Q}_t = [\boldsymbol{q}_t^1, \boldsymbol{q}_t^2, \dots, \boldsymbol{q}_t^N] $, where each $ \boldsymbol{q}_t^{i} \in \mathbb{R}^{d} $. For frame-level processing, VGG-16 is employed to extract global features, ensuring consistency across spatial and object-based representations.

\subsection{Efficient Multiscale Spatial Aggregation (EMSA)}
The EMSA module is designed to reduce computational costs while preserving spatial feature richness, a critical aspect for early accident anticipation. Traditional convolutional approaches often struggle with scalability, as the number of parameters grows quadratically with kernel size and channel dimensions. EMSA mitigates this issue by introducing \textit{feature grouping}, which partitions the multi-scale input tensor $ \mathbf{O}_t $ into $ G $ smaller sub-groups along the channel axis. By recalibrating spatial attention weights within each sub-group, EMSA significantly reduces the parameter count to $ \frac{1}{G} $ of the traditional approach. This design enhances computational efficiency while strengthening the model’s ability to focus on localized regions, which is crucial for capturing subtle spatial cues.

Effective modeling of global and local spatial information in EMSA utilizes a multiscale architecture incorporating two complementary representations: \textit{coarse-grained} and \textit{fine-grained feature maps}. Coarse-grained maps derive from 2D global pooling operations that condense feature information across expanded receptive fields. The term ``2D'' specifically refers to spatial dimension reduction along both height ($H$) and width ($W$), where each channel's activation map is compressed into a single scalar value through averaging. Such spatial aggregation captures inter-object relationships while maintaining computational efficiency, an approach particularly advantageous when processing high-resolution inputs. Fine-grained maps retain pixel-level precision while hierarchically integrating multi-scale contextual information. These dual representations synergistically enhance accident anticipation capabilities, with fine-grained features emphasizing the exact spatial configurations required for precise accident anticipation.

EMSA processes the multi-scale input tensor $ \mathbf{O}_t \in \mathbb{R}^{b \times C \times H \times W} $, where $ b $ is the batch size, $ C $ is the number of input channels, and $ H, W $ are the spatial dimensions. The tensor is divided into $ G $ sub-groups along the channel dimension, reshaped as $ \mathbf{O}_t^{(g)} \in \mathbb{R}^{b \times (C/G) \times H \times W} $ for each group $ g $, enabling efficient attention modeling. Two parallel convolutional branches—$ 1 \times 1 $ and $ 3 \times 3 $—process the grouped feature maps. The $ 1 \times 1 $ branch extracts high-level global spatial relationships, while the $ 3 \times 3 $ branch captures localized spatial details through an expanded receptive field. The 2D global pooling operation used in both branches is expressed as:
\begin{equation}
    z_{t,c}^{(g)} = \frac{1}{H \times W} \sum_{j=1}^{H} \sum_{i=1}^{W} o_{t,c}^{(g)}(i, j)
\end{equation}
where $ z_{t,c}^{(g)} $ represents the pooled feature value for channel $ c $ in group $ g $ at time step $ t $. The pooled representation is followed by a shared $ 1 \times 1 $ convolution and a non-linear sigmoid activation, approximating a 2D binomial distribution. Attention maps produced by parallel branches undergo matrix dot-product fusion to facilitate cross-channel communication, thereby generating comprehensive attention maps that encode multi-scale spatial dependencies. Subsequent sigmoid-based refinement enhances these features by simultaneously capturing pixel-wise correlations and global contextual patterns. The resulting output tensor $ \mathbf{O}_t^{\prime} \in \mathbb{R}^{b \times 2d \times H \times W} $ doubles the channel dimension through feature concatenation—where the original $ C=2d $ channels from multi-scale inputs are expanded by aggregating complementary coarse and fine-grained representations.

\subsection{Memory Attention Aggregation (MAA)}
Self-attention mechanisms \citep{zhang2024self, adewopo2023review, xie2024real} are widely recognized for their ability to enhance the representational capacity of temporal feature modules. However, their computational requirements scale quadratically with the input size, as they involve processing and storing relationships across all positional embeddings. Such an approach presents significant challenges for tasks involving long image sequences, particularly in resource-constrained environments. In accident anticipation scenarios requiring continuous frame sequence analysis for temporal correlation extraction, such computational complexity directly compromises real-time operational feasibility. Moreover, conventional self-attention mechanisms \citep{geng2024stgaformer, duan2022fdsa, zhang2023traffic} often focus exclusively on intra-sample positional relationships, neglecting inter-sample correlations. This limitation reduces generalization performance across heterogeneous datasets, thereby impacting the robustness of anticipation.

To address these challenges, we propose the \textit{Memory Attention Aggregation (MAA)} module, designed to capture temporal dependencies across sequences efficiently. At the core of MAA are two \textit{memory units} with significantly reduced dimensionality compared to the input features. These units are arranged in parallel within independent linear layer structures, enabling the extraction of abstract yet informative representations without incurring prohibitive computational costs. By alleviating the limitations of traditional self-attention mechanisms, MAA enhances predictive accuracy in accident anticipation tasks while maintaining computational efficiency.

The MAA module first projects the multi-scale input tensor $ \mathbf{O}_t \in \mathbb{R}^{b \times C \times H \times W} $ into a latent attention space via linear transformations, generating both an attention map $ A_t $ and a memory-enhanced representation $ H_t' $, as follows:
\begin{align}
    \mathbf{M}_t &= \mathbf{O}_t \mathbf{W}_{\mathrm{mk}}, \\
    \mathbf{A}_t &= \operatorname{softmax}(\mathbf{M}_t), \\
    \mathbf{H}^{\prime}_t &= \mathbf{A}_t \mathbf{M}_t
\end{align}
where $ \mathbf{W}_{\mathrm{mk}} \in \mathbb{R}^{C \times S} $ denotes a trainable projection matrix mapping features into a memory key subspace of dimensionality $ S $. The memory tensor $ \mathbf{M}_t \in \mathbb{R}^{b \times (H \times W) \times S} $ dynamically encodes inter-position correlations, while the output $ \mathbf{H}_t' \in \mathbb{R}^{b \times (H \times W) \times S} $ synthesizes these dependencies through attention-guided aggregation.

The refined attention maps subsequently undergo dimensional reconstruction through learnable linear transformations, bridging the compressed memory subspace back to the original feature dimensions. This restoration process yields the memory-augmented representation $ \mathbf{O}_{\mathrm{t}}^{\text{mem}} $, formally expressed as:
\begin{equation}
    \mathbf{O}_{\mathrm{t}}^{\text{mem}} = \mathbf{A}_t \mathbf{W}_{\mathrm{mv}}
\end{equation}
where $\mathbf{W}_{\mathrm{mv}} \in \mathbb{R}^{S \times C} $. The projection matrix $ \mathbf{W}_{\mathrm{mv}} $ mediates feature reconstruction while preserving the original channel structure from the multi-scale concatenation process. Crucially, this memory-enriched tensor encapsulates both spatially attended patterns and temporally distilled dependencies through its hybrid composition - maintaining the critical $ C $-dimensional feature structure while embedding latent spatiotemporal relationships essential for reliable accident anticipation.

To capture sequential dynamics, temporal attention weights $ \boldsymbol{\beta}_t $ are computed through non-linear transformations:
\begin{equation}
    \boldsymbol{\beta}_{t} = \gamma\left(\mathbf{W}_{\mathrm{ta}} \tanh \left(\mathbf{H}^{\prime}_{t}\right)\right) \mathbf{O}_{\mathrm{t}}^{\text{mem}}
\end{equation}
where $ \mathbf{W}_{\mathrm{ta}} \in \mathbb{R}^{C \times C} $ parameterizes the temporal interaction space, and $ \gamma(\cdot) $ denotes an element-wise activation function. The final temporally aggregated features are then derived via convolutional fusion:
\begin{equation}
    \boldsymbol{h}_{t}^{\prime} = \left\langle\boldsymbol{\beta}_{t}, (\mathbf{O}_t)\right\rangle_r
\end{equation}
where $ \langle, \rangle_r $ represents a depthwise convolution operation that aligns temporal correlations across sequential states. These synthesized features $ \boldsymbol{h}_{t}^{\prime} \in \mathbb{R}^{b \times C} $ encapsulate motion-critical relationships essential for reliable accident anticipation.

\subsection{Auxiliary Self-Attention Aggregation (AAA)} 
Traffic accidents often manifest through subtle and complex indicators across consecutive video frames, such as vehicle deceleration or anomalous motion patterns that typically require extended temporal observation to detect. Traditional approaches \citep{liang2023memory, santhosh2020anomaly, rezaee2024survey} frequently struggle to capture latent dependencies between temporally distant frames, consequently limiting their accident anticipation accuracy. To address this limitation, the \textit{Auxiliary Self-Attention Aggregation (AAA)} module analyzes frame-to-frame relationships by adaptively assigning weights according to contextual relevance. Through selective amplification of accident-related features and suppression of irrelevant signals, AAA effectively integrates multi-scale temporal patterns essential for reliable early warning.

Although large model parameters improve anticipation accuracy, their high computational costs and compromised real-time performance create implementation barriers for accident anticipation systems in real-world scenarios. This issue becomes particularly critical in accident anticipation where high-resolution spatial-temporal feature preservation is paramount. To maintain computational tractability without compromising feature integrity, the AAA module adopts \textit{depthwise separable convolutions}. These operations decouple standard convolutions into two stages—channel-wise spatial filtering followed by linear feature combination—establishing a lightweight structure that achieves accelerated inference speeds while maintaining anticipation fidelity \citep{khalifa2024real}.

While bottleneck layers \citep{lin2024robust, gupta2021deep, latif2023deep} successfully mitigate vanishing gradient issues via identity shortcut connections, their architectural reliance on aggressive dimensionality reduction—characterized by sequential compression-restoration operations—risks gradual spatial information erosion. Depthwise separable convolutions \citep{khalifa2024real, shen2021vehicle} alternatively decompose standard convolution into two complementary phases: spatial filtering through \textit{depthwise convolution} maintaining channel integrity, followed by cross-channel fusion via \textit{pointwise convolution}, attaining enhanced parameter efficiency compared to conventional approaches \citep{khalifa2024real} without altering original feature dimensions. The framework’s dimension-preserving design critically retains fine-grained spatial semantics essential for early accident anticipation. By maintaining uniform feature resolution through cross-dimensional interaction layers, it prevents information erosion in bottleneck structures—a limitation inherent to compression-based paradigms where aggressive dimensionality reduction disproportionately attenuates discriminative spatiotemporal cues during feature abstraction.

The module's self-attention mechanism processes the multi-scale input tensor $\mathbf{O}_t \in \mathbb{R}^{b \times C \times H \times W}$ through spatial-temporal interaction modeling, defined by $\mathbf{F}_v^{\prime} = \gamma \langle \mathbf{O}_t^{\mathsf{T}}, \mathbf{O}_t \rangle_r$ where $\mathbf{O}_t^{\mathsf{T}} \in \mathbb{R}^{C \times b \times H \times W}$ denotes channel-transposed inputs, $\langle \cdot, \cdot \rangle_r$ indicates depthwise convolution with $r$-sized receptive fields, and $\gamma$ implements nonlinear activation. Parameters $\mathbf{W}_{\mathrm{aaa}} \in \mathbb{R}^{C \times d}$ are optimized through back-propagation of a composite loss function combining frame-level accident scores with attention entropy regularization, enabling adaptive focus on critical temporal windows across diverse accident scenarios while maintaining original spatial resolution.

The final aggregated representation $ \boldsymbol{Z}_v^{\prime} \in \mathbb{R}^d $ synthesizes these contextualized features through two steps:
\begin{equation}
\boldsymbol{Z}_v = \mathcal{G}(\mathbf{F}_v^{\prime}) \mathbf{W}_{\mathrm{aaa}} 
\end{equation}
where $\mathcal{G}$ denotes global pooling. Subsequent processing involves depthwise separable convolution and two factorized fully-connected layers with shared parameters $ \mathbf{B}_v = \{ \mathbf{B}_{v0}, \mathbf{B}_{v1} \} \subset \mathbb{R}^{d \times d} $, resulting in:
\begin{equation}
Z_v^{\prime} = \operatorname{Softmax}\left( \phi\left( \phi\left( Z_v^{\prime} \mathbf{B}_{v0} \right) \mathbf{B}_{v1} \right) \right)
\end{equation}
where $ \phi(\cdot) $ denotes swish activation function. The refined feature $\boldsymbol{Z}_v^{\prime}$ preserves critical spatial details through swish-activated transformation while ensuring computational efficiency via parameter reuse across temporal scales.

\subsection{Flamingo Alert-Assisted System (FAA)}
Recent advancements in autonomous driving systems have increasingly leveraged natural language descriptions to enhance scene understanding, situation awareness, and human-machine interaction \citep{zhou2024vision, atakishiyev2024explainable, smith2022natural}. Designed for lightweight operation, \textit{Flamingo Alert-Assisted System } complements the accident anticipation pipeline by generating context-aware natural language notification of accident and converting them into actionable verbal alerts.

The Flamingo architecture \citep{chowdhury2023flamingo} was selected for traffic accident anticipation due to its effective integration of multimodal processing, computational efficiency, and operational flexibility -- essential characteristics for real-time analysis in dynamic driving environments. Compared to traditional vision-language models like GPT-2 \citep{lee2024fractal, qu2020text} and GPT-3 \citep{hinton2023persuasive, gan2024large} that require computationally expensive multimodal training, Flamingo achieves efficient visual-text integration through optimized architecture design, as demonstrated in Tables~\ref{tab:model_comparison}. This design enables rapid analysis of traffic scenarios with low processing latency, producing context-aware safety alerts crucial for autonomous driving systems. The framework's parameter-efficient design maintains strong pattern recognition capabilities for both visual and textual data while delivering consistent performance across diverse traffic conditions, including congested urban intersections and high-speed highways \citep{bathla2022autonomous, johnson2021modular}.

The FAA framework's operational pipeline initiates with dashcam video frame processing through a Cascade R-CNN detector, extracting spatial-temporal features including bounding box coordinates and accident probability estimates. These features undergo parallel processing: (i) CLIP-embedded semantic recognition identifying accident precursors through contrastive visual-text alignment \citep{radford2021learning}, and (ii) Perceiver Resampler tokenization transforming variable-resolution frames into fixed-dimensional visual embeddings. A dedicated linguistic interface module processes textual prompts using GPT-3.5's tokenization schema \citep{openai2023gpt}, with syntactic normalization ensuring compatibility with Flamingo's cross-attention mechanisms \citep{alayrac2022flamingo}. The co-evolution of linguistic tokens (from GPT-3.5) and visual embeddings occurs through Flamingo's gated cross-attention layers, where adaptive projection matrices mediate between the distinct token spaces while preserving modality-specific features. This hybrid architecture implements context-sensitive fusion where GPT-3.5-derived linguistic tokens dynamically gate visual feature integration through learnable attention masks. The framework maintains Flamingo's core capability of processing interleaved visual-text sequences while augmenting its linguistic foundation with GPT-3.5's semantic comprehension. During decoding, the architecture employs constrained beam search with GPT-3.5's vocabulary priors to generate safety-critical descriptions containing accident probabilities, object detections, and spatial relationships. These outputs interface with a text-to-speech module through latency-optimized API endpoints, achieving real-time alert generation through computational optimizations in the system architecture design.

\begin{table}[ht]
\centering
\caption{Comparative Analysis of Key Features in Traffic Accident Anticipation: Flamingo vs. GPT-2 vs. GPT-3}
\label{tab:model_comparison}
\resizebox{0.9\textwidth}{!}{
\begin{tabular}{c|c|c|c}
\hline
\textbf{Feature} & \textbf{Flamingo} & \textbf{GPT-2} \citep{lee2024fractal} & \textbf{GPT-3} \citep{hinton2023persuasive} \\
\hline
Cross-Modal Understanding & \textbf{High} & Low & Mid \\
Real-Time Inference & \textbf{High} & Low & Mid \\
Computational Efficiency & \textbf{High} & Low & Low \\
Traffic Scenario Adaptability & \textbf{High} & Low & Mid \\
Scene Interpretation & \textbf{High} & Low & Mid \\
Autonomous System Integration & \textbf{High} & Low & Mid \\
Alert Responsiveness & \textbf{High} & Low & Mid \\
\hline
\end{tabular}
}
\vspace{3mm}
\end{table}

\subsection{Training}
\label{sec:training}

The LATTE framework employs a dual supervision strategy that jointly optimizes temporal localization precision and holistic video understanding through complementary loss formulations. The training process operates at two temporal granularities to address both frame-level event anticipation and video-level semantic consistency. The frame-level supervision enforces temporally aware accident localization by imposing exponentially increasing penalties as predictions approach critical events. For each video $v$, let $\mathbf{I}_v^{\text{acc}} \in \{0,1\}$ denote the accident occurrence indicator and $p_t^{(v)} \in [0,1]$ represent the predicted accident probability at frame $t$. The temporal weighting function $\omega(t,\tau) = \exp(\beta(\tau - t)_+)$ introduces temporal urgency awareness through an exponential decay mechanism, where $\tau$ denotes the ground truth accident onset time and $\beta \in \mathbb{R}^+$ controls the exponential decay rate:

\begin{equation}
\mathcal{L}_{\text{frame}} = -\sum_{v=1}^N \Bigg[ \mathbf{I}_v^{\text{acc}} \sum_{t=1}^T \omega(t,\tau_v) \log p_t^{(v)} + (1 - \mathbf{I}_v^{\text{acc}}) \sum_{t=1}^T \log(1 - p_t^{(v)}) \Bigg]
\label{eq:frame_loss}
\end{equation}
where $(x)_+ = \max(x,0)$ ensures non-negative temporal intervals. The decay rate parameter $\beta$ governs how rapidly the loss weight increases as the prediction approaches the accident moment - larger $\beta$ values create sharper exponential growth in penalty weights near $\tau$. The video-level loss operates on the temporal maximum pooling output $p_{\text{vid}}^{(v)} = \max_t p_t^{(v)}$, enforcing global semantic alignment across the entire video sequence:

\begin{equation}
\mathcal{L}_{\text{video}} = -\sum_{v=1}^N \Big[ \mathbf{I}_v^{\text{acc}} \log p_{\text{vid}}^{(v)} + (1 - \mathbf{I}_v^{\text{acc}}) \log(1 - p_{\text{vid}}^{(v)}) \Big]
\label{eq:video_loss}
\end{equation}

The composite objective function combines these components through adaptive weighting:

\begin{equation}
\mathcal{L}_{\text{total}} = \mathcal{L}_{\text{frame}} + \lambda \mathcal{L}_{\text{video}}
\label{eq:total_loss}
\end{equation}
where $\lambda$ balances the relative importance of global video classification. This dual formulation ensures simultaneous optimization of precise temporal localization (via $\mathcal{L}_{\text{frame}}$) and accurate video-level accident anticipation (through $\mathcal{L}_{\text{video}}$). The exponential weighting in Equation~\ref{eq:frame_loss} creates temporal urgency pressure during gradient updates through the $\beta$-controlled decay mechanism, while the max-pooling operation in Equation~\ref{eq:video_loss} encourages at least one high-confidence prediction per positive video sequence.

\section{Experiment}
\subsection{ Datasets }
\begin{figure}
    \centering
    \includegraphics[width=0.5\linewidth]{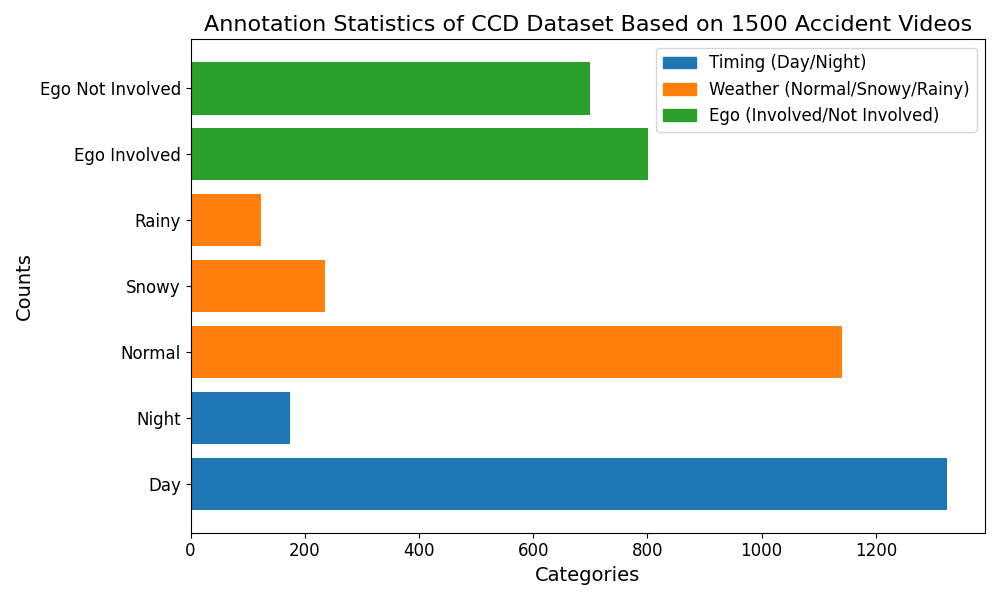}
    \caption{Annotation Statistics of CCD Dataset. The histogram emphasizes environmental condition variability (weather patterns and illumination states), ego-vehicle engagement dynamics, and scenario complexity across 4,500 annotated clips. The stratified training-test partition (4:1 ratio) ensures robust evaluation of accident anticipation systems, which enables precise modeling of traffic interactions across heterogeneous driving contexts, significantly advancing proactive accident anticipation system development through scenario-aware learning paradigms.}
    \label{fig:enter-label3}
\end{figure}

\begin{figure}
    \centering
    \includegraphics[width=0.75\linewidth]{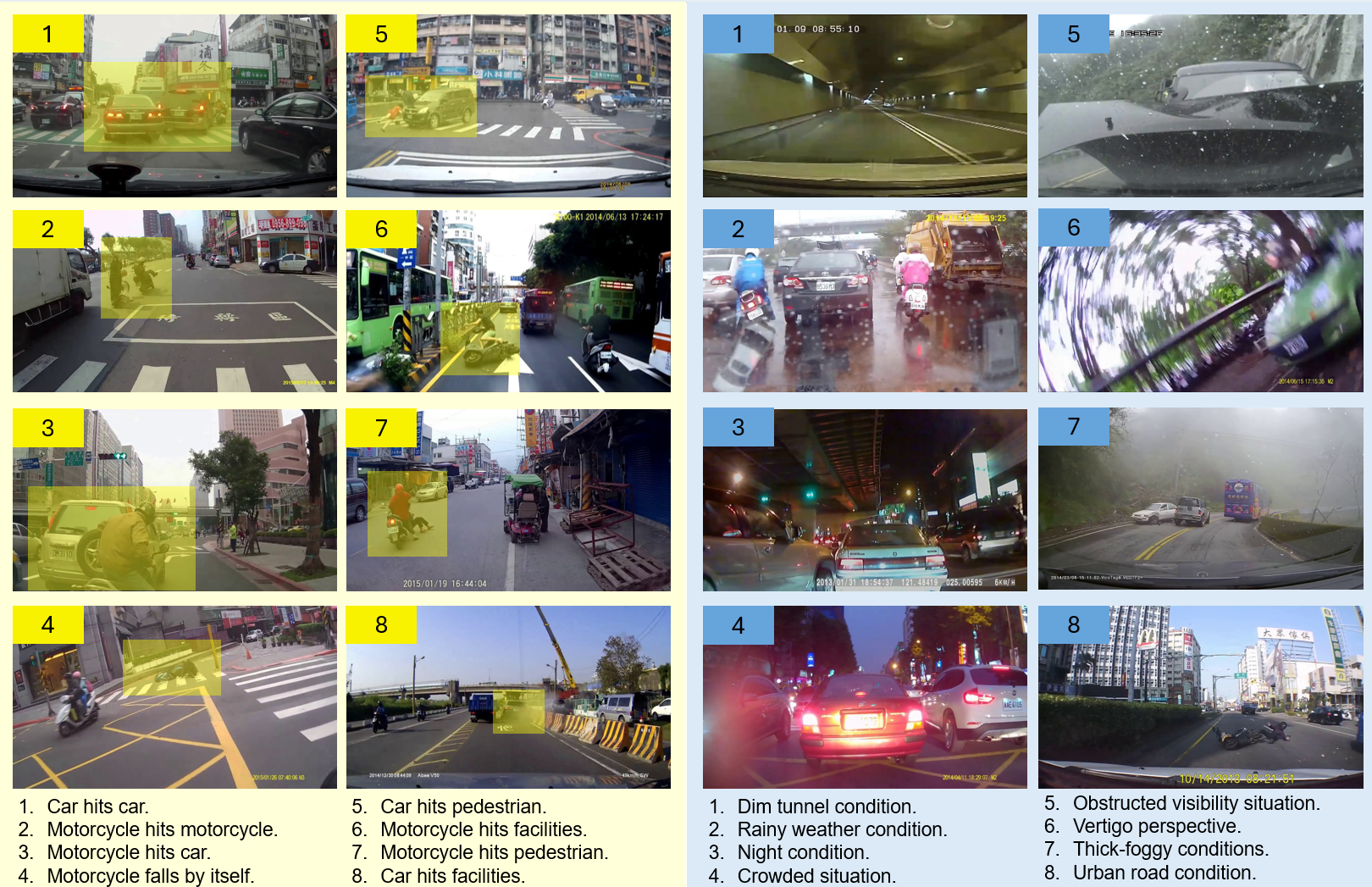}
    \caption{Visualization of multi-category accident instances in the DAD dataset, showcasing:Diverse detected traffic participants (marked by the yellow box) and accident types; (b) Scenario variations encompassing meteorological conditions (rain/snow/fog), illumination levels (daytime/night), and perspective configurations.}
    \label{fig:enter-label4}
\end{figure}

We evaluate LATTE using three publicly available datasets that focus on accident anticipation:
\begin{itemize}
    \item \textbf{CCD:} The Car Crash Dataset (CCD) \citep{bao2020uncertainty} provides detailed annotations of environmental factors, ego-vehicle involvement, accident participants, and causal mechanisms. Comprising 1,500 positive clips (accident-containing) and 3,000 negative clips (accident-free), the dataset is partitioned into 3,600 training clips and 900 testing clips, with each clip containing 50 frames spanning 5 seconds. Figure~\ref{fig:enter-label3} visually demonstrates the dataset's scenario diversity, highlighting its capacity to model real-world traffic dynamics and enhance accident anticipation frameworks.
    \item \textbf{DAD:} The Dashcam Accident Dataset (DAD) \citep{chan2017anticipating} contains 720p-resolution dashcam footage collected across six major Taiwanese cities. Its 620 positive clips and 1,130 negative clips are divided into 1,284 training clips and 466 testing clips, each comprising 100 frames over 5 seconds. The dataset covers multiple accident types including car-motorcycle, car-to-car, and motorcycle-to-motorcycle incidents, with representative samples shown in Figure~\ref{fig:enter-label4}. These exemplify the dataset's effectiveness in training and evaluating accident anticipation models under diverse conditions, including edge cases.
    \item \textbf{A3D:} The AnAn Accident Detection Dataset (A3D) \citep{yao2019unsupervised} documents abnormal road events across East Asian urban environments. With 1,087 positive clips and 114 negative clips divided into 961 training clips and 240 testing clips, each 5-second sequence contains 100 frames. A3D maintains temporal and structural configurations identical to DAD for traffic accident anticipation research.
\end{itemize}

\subsection{ Metrics }
To assess the performance of LATTE, we use three key metrics, each capturing a unique aspect of accident anticipation:
   \begin{itemize}
    \item \textbf{Average Precision (AP):} AP is computed by integrating the precision values across varying recall levels, providing a balanced measure of the trade-off between precision and recall. Precision (\(P\)) and recall (\(R\)) are defined as:
    \begin{equation}
    R = \frac{TP}{TP + FN}, \quad P = \frac{TP}{TP + FP}
    \end{equation}
    where \(TP\), \(FP\), and \(FN\) represent the true positives, false positives, and false negatives, respectively. AP quantifies the area under the precision-recall (PR) curve as:
    \begin{equation}
    \text{AP} = \int_0^1 P(r) \, dr
    \end{equation}
    where \(P(r)\) denotes precision as a function of recall \(r\). A higher AP indicates strong classification performance with minimal false positives and negatives. 

    \item \textbf{Time-To-Accident at R80 (TTA@R80):} TTA@R80 measures how early an accident can be anticipated when the model achieves a recall rate of 80\%. It is computed as:
    \begin{equation}
    \text{TTA@R80} = \text{Average }(t_a - t_p) \quad \text{for } R \geq 0.80
    \end{equation}
    where \(t_p\) and \(t_a\) represent the predicted and actual times of the accident, respectively. A higher TTA@R80 indicates better early warning capabilities under high-recall constraints.
    
    \item \textbf{Mean Time-To-Accident (mTTA):} mTTA represents the average time-to-accident for all positive samples. For \(N\) positive samples with individual TTA values \(TTA_i\), mTTA is computed as:
    \begin{equation}
    mTTA = \frac{1}{N} \sum_{i=1}^N TTA_i
    \end{equation}
    A higher mTTA reflects improved foresight in accident anticipation.

    \item \textbf{Floating Point Operations (FLOPs):} FLOPs quantify the arithmetic operations required for a single forward pass through the model, serving as an indicator of computational complexity. Models with reduced FLOPs demonstrate enhanced operational efficiency, thereby increasing their suitability for deployment in resource-constrained environments.
    
    \item \textbf{Parameter Count (Params):} Params measure the total number of learnable weights in the model, representing its capacity. While higher Params can enhance representational power, they may also increase memory usage and risk overfitting.
\end{itemize}

\subsection{Implementation Details}
All experiments were conducted on an NVIDIA GeForce RTX 4080 GPU. The DAD dataset was preprocessed using VGG-16, with the hidden state dimension set to 512. The model was implemented in PyTorch 3.7, trained for 15 epochs with a learning rate of \(1 \times 10^{-3}\) and a batch size of 10.

\subsection{Comparison to State-of-the-art (SOTA) }
Table~\ref{tab:my_label1} summarizes LATTE's performance across the DAD, CCD, and A3D datasets. The framework consistently achieves robust benchmarking results in diverse evaluation scenarios. LATTE achieves equivalent anticipation accuracy to state-of-the-art methods on CCD and A3D benchmarks, while demonstrating superior cross-domain generalization on DAD with 29.7\% AP elevation and 18.2\% mTTA improvement, validating its scenario-agnostic reliability through rigorous leave-one-dataset-out validation protocols. These achievements are attained alongside substantially reduced computational demands, confirming practical deployability in resource-limited environments.

While LATTE demonstrates significant performance improvements over existing models on the DAD dataset, its accuracy metrics remain slightly below those achieved on the CCD and A3D benchmarks. The observed performance difference originates from the DAD dataset's inherent complexity and scenario diversity, as illustrated in Fig.~\ref{fig:enter-label4}. In contrast to the controlled accident simulations characteristic of CCD and A3D, the DAD benchmark incorporates more challenging environmental variables including variable illumination, meteorological conditions, and roadway geometries that increase accident anticipation difficulty. These compounding factors amplify data distribution heterogeneity, leading to comparatively reduced DAD performance despite LATTE's demonstrated excellence across alternative benchmarks. Notably, LATTE maintains a substantial AP improvement over baseline models on DAD, confirming its operational robustness and scenario adaptability in complex real-world accident contexts.

    \begin{table}
    \centering
    \caption{Comparison of models balancing AP and mTTA across three datasets. The top and second-best performances in each category are marked in \textbf{bold} and \underline{underlined}, respectively. Missing values are indicated by a dash (``-'').}
    \label{tab:my_label1}
    \resizebox{1\textwidth}{!}{
    \textbf{}
    \begin{tabular}{c|c|c|c|c|c|c} 
        \hline
        & \multicolumn{2}{c|}{DAD} & \multicolumn{2}{c|}{CCD} & \multicolumn{2}{c}{A3D} \\ 
        \hline
        Models & AP (\%) & mTTA (s) & AP (\%) & mTTA (s) & AP (\%) & mTTA (s) \\ 
        \hline
        DSA \citep{chan2017anticipating} & 48.1 & 1.34 & 98.7 & 3.08 & 92.3 & 2.95 \\ 
        ACRA \citep{zeng2017agent} & 51.4 & 3.01 & 98.9 & 3.32 & - & - \\ 
        AdaLEA \citep{suzuki2018anticipating} & 52.3 & 3.43 & 99.2 & 3.45 & 92.9 & {3.16} \\ 
        Ustring \citep{bao2020uncertainty} & 53.7 & 3.53 & {99.5} & 3.45 & 92.9 & {3.16} \\ 
        DSTA \citep{karim2022dynamic} & 56.1 & {3.66} & \underline{99.6} & {3.87} & 93.5 & 2.87 \\
        GSC \citep{wang2023gsc} & {60.4} & 2.55 & 99.4 & 3.68 & {94.9} & 2.62 \\ 
        CRASH \citep{liao2024crash} & {65.3} & 3.05 & \underline{99.6} & \textbf{4.91} & \underline{96.0} & \textbf{4.92}
        \\
        W3AL \citep{liao2024and} & \underline{69.2} & \underline{4.26} & \textbf{99.7} & {3.93} & \textbf{96.4} & {3.48}\\
        \textbf{LATTE}& \textbf{89.74} & \textbf{4.49} & 98.77 & \underline{4.53} & 92.46 & \underline{4.52} \\ 
        \hline
    \end{tabular}
    }
\end{table}

Furthermore, LATTE’s performance on the DAD dataset is evaluated by comparing its best AP, mTTA, and TTA@R80 metrics against those of other models, as presented in Table~\ref{tab:my_label2}. The results indicate that LATTE consistently outperforms existing approaches, achieving an AP of 89.47\%——26.7\% higher than the second-best model CRASH—thereby underscoring its advanced accident anticipation capabilities and strong potential for early warning applications in autonomous vehicles. In contrast, convolutional architectures such as those in DSA and Ustring effectively capture local spatial-temporal relationships but struggle with long-range dependencies and global context due to their fixed kernel structures. Likewise, graph-based methods like GSC and W3AL center on localized spatial-temporal dynamics but often face scalability challenges in complex traffic scenarios, limiting their generalizability. 

Notably, LATTE significantly outperforms the runner-up method W3AL with 5.4\% improvement in mTTA. While achieving state-of-the-art TTA@R80 performance at 4.04 seconds, LATTE simultaneously maintains second-best overall ranking across all evaluation metrics. These metrics collectively validate the framework's capacity for accurate accident anticipation and extended early-warning time windows, directly contributing to accident risk reduction and improved roadway safety through proactive hazard anticipation.
    
    \begin{table}
    \centering
    \caption{Comparison of models for the highest AP, mTTA, and TTA@R80 on the DAD dataset. The top and second-best performances in each category are marked in \textbf{bold} and \underline{underlined}, respectively.}
    \label{tab:my_label2}
    \begin{tabular}{c|c|c|c}
        \hline
        Models & AP (\%) & mTTA (s) & TTA@R80 (s) \\
        \hline
        Ustring \citep{bao2020uncertainty} & 68.40 & 1.63 & 2.18 \\
        XAI-Accident \citep{monjurul2021towards} & 64.32 & {1.80} & 0.68 \\
        DSTA \citep{karim2022dynamic} & 66.70 & 1.52 & {2.39} \\
        GSC \citep{wang2023gsc} & {68.90} & 1.33 & 2.14 \\
        CRASH \citep{liao2024crash} & \underline{70.86} & 1.91 & 2.20 \\
        W3AL \citep{liao2024and} & 69.20 & \underline{4.26} & \textbf{4.33} \\
        \textbf{LATTE}& \textbf{89.74} & \textbf{4.49} & \underline{4.04} \\
        \hline
    \end{tabular}
\end{table}

The primary objectives of LATTE are a lightweight design and high computational efficiency, as demonstrated by comparing its FLOPs and Params to other SOTA models on the DAD dataset (Table~\ref{tab:my_label3}). The results show that LATTE substantially outperforms existing models in terms of computational efficiency. Specifically, it reduces FLOPs by about 93.14\% relative to the second-best model (DSTA) and by over 5,917 times compared to the largest model (UniFormerV2). LATTE also lowers Params by 31.58\% compared to DSTA, and by even greater margins compared to other SOTA approaches. LATTE's attention-based framework effectively captures both local and global spatiotemporal features through dynamic processing, enabling robust handling of complex accident scenarios while maintaining scalability. Unlike computationally intensive temporal models such as AdaLEA and DSTA, LATTE minimizes redundancy via lightweight attention mechanisms, thereby substantially reducing computational overhead (Table~\ref{tab:my_label3}). Furthermore, FAA improves human-vehicle trust through context-aware textual feedback which is often neglected in earlier models. These advancements ensure not only improved predictive accuracy but also greater practical efficiency, making LATTE particularly suitable for real-time deployment in resource-constrained environments.

Through architectural refinements and optimized computational strategies, LATTE eliminates non-essential operations while focusing on critical processes, resulting in enhanced AP and mTTA performance compared to other models (Table~\ref{tab:my_label1}). This computational efficiency enables faster inference speeds, reduced energy consumption, and compatibility with resource-limited hardware. Additionally, the reduced parameter count decreases LATTE's memory footprint, enabling faster experimental iterations and more efficient development cycles.

\begin{table}
\centering
\caption{Comparison of models in efficiency. FLOPs denotes floating point operations and Params means parameter count. The top and second-best performances in each category are marked in \textbf{bold} and \underline{underlined}, respectively.}
\label{tab:my_label3}
    \begin{tabular}{c|c|c}
         \hline
         Models&  FLOPs (M)&  Params (M)\\
         \hline
         UniFormerV2 \citep{li2022uniformerv2}&  3600000.00&  115.00\\
         VideoSwin \citep{liu2022video}&  282000.00&  88.10\\
         MViTv2 \citep{fan2021multiscale}&  206000.00&  51.00\\
         DSTA \citep{karim2022dynamic}&  \underline{8868.0}0&  \underline{4.56}\\
         \textbf{LATTE}& \textbf{608.34}&\textbf{3.12}\\
         \hline
    \end{tabular}
\end{table}

\subsection{Ablation Studies}
 
Investigating the contributions of individual modules in the LATTE model, we performed an ablation study on the DAD dataset with emphasis on three core components: EMSA, MAA, and AAA. As evidenced in Table~\ref{tab:my_label4}, the results quantify each module's influence on performance metrics and computational efficiency, providing critical insights into the model's scalability and real-world deployment potential. Although high accuracy (AP, mTTA) remains essential for reliable accident anticipation, auxiliary metrics including Frames Per Second (FPS) and FLOPs expose underlying computational requirements. In resource-constrained scenarios, models attaining superior AP or mTTA at the expense of excessive FLOPs often become operationally infeasible. Conversely, exclusive prioritization of computational efficiency may degrade anticipation reliability. LATTE resolves this dichotomy through streamlined attention mechanisms that minimize FLOPs while preserving both high FPS and accuracy thresholds. This equilibrium supports precise accident identification and resource-efficient execution on edge devices, guaranteeing dependable real-time performance. We consequently propose three complementary metrics—FPS, FLOPs, and Params—for multidimensional evaluation of accuracy, computational expenditure, and scalability.

To assess EMSA impact, we compare Model A (excluding EMSA but including MAA and AAA) with the original model (including all modules). Model A results in a significant decrease in AP from 89.74\% to 85.60\%. Additionally, FPS drops from 1508.47 to 665.79, indicating a substantial reduction in frame processing efficiency. The lower FPS suggests that without EMSA, the model struggles with frame selection efficiency, leading to slower processing times. Its exclusion may hinder the model's ability to process dense traffic scenarios or environments with complex spatial layouts, such as urban intersections. Interestingly, Model A has lower FLOPs (298.39) compared to the original model (608.34), and slightly fewer Params (3.05 vs. 3.12). However, despite the higher computational cost in the original model, the FPS is more than doubled, and the AP is significantly improved, which indicates that EMSA is critical for frame selection efficiency, enhancing both accuracy and processing speed. The significant drop in AP when EMSA is excluded can be attributed to its role in aggregating spatial features, which enables the model to capture fine-grained spatial information that is vital for precise accident anticipation. Without EMSA, the model loses the capability to focus on key spatial regions, thus leading to a drop in accuracy. By preserving spatial feature and optimizing computations, EMSA allows the model to process frames more rapidly, which is essential for real-time accident anticipation.

The critical role of MAA emerges through comparative analysis of Model B (MAA-excluded configuration with retained EMSA and AAA) against the complete architecture. Performance metrics reveal a substantial decrease in AP to 84.78\% from 89.74\% in the original model. The mTTA also slightly decreases from 4.49 seconds to 4.40 seconds, and TTA@R80 increases from 4.04 seconds to 4.65 seconds, indicating less effective early anticipation capabilities. The FPS in Model B is 1484.04, slightly lower than the original model's 1508.47. The FLOPs exhibit a marginal reduction (607.73 vs. 608.34), while the Params remain effectively constant at 3.12. The minimal differences in computational costs suggest that MAA plays a vital role in modeling temporal dependencies and enables the extraction of informative representations without incurring significant computational costs, making it indispensable for handling long sequences typical in dashcam videos. By modeling long-term temporal dependencies, MAA is sensitive to scenarios where accident cues unfold over an extended period, such as gradual lane drifts or prolonged braking patterns. The exclusion of MAA results in a loss of critical temporal context, as it captures long-range dependencies that are essential for making accurate anticipations in sequences with varying dynamics. The drop in AP reflects the model's inability to properly track and anticipate accidents in longer temporal sequences, which is crucial for real-time accident anticipation.

To understand the impact of AAA, we analyze the performance of Model C, which excludes AAA but includes EMSA and MAA, in comparison to the original model that integrates all modules. Excluding AAA  leads to a reduction in AP from 89.74\% to 87.33\%. the mTTA drops from 4.49 seconds to 3.38 seconds, and TTA@R80 decreases from 4.04 seconds to 3.68 seconds, indicating less effective anticipation of accidents. What's more, Model C has a higher FPS (1671.90) compared to the original model (1508.47), and slightly lower FLOPs (576.43 vs. 608.34). The decrease in computational complexity and increase in FPS are due to the exclusion of AAA, which, while computationally demanding, contributes to capturing extended temporal dependencies by focusing on contextual relevance. Despite the increase in FPS, the drop in AP can be attributed to the removal of AAA’s ability to prioritize contextual features, which allows the model to focus on accident-relevant patterns over longer time periods. AAA enhances contextual understanding, which is crucial for interactions between multiple agents in diverse traffic conditions. Without AAA, the model becomes less capable of capturing important contextual cues, leading to a decline in predictive accuracy.By effectively prioritizing accident-related features, AAA significantly enhances the model's predictive performance.

    \begin{table}
        \centering
        \caption{Ablation results for the DAD dataset. EMSA, MAA, and AAA denote Efficient Multiscale Spatial Aggregation, Memory Attention Aggregation, and Auxiliary Self-Attention Aggregation, respectively. The top and second-best performances in each category are marked in \textbf{bold} and \underline{underlined}, respectively.}
        \label{tab:my_label4}
        \begin{tabular}{c|c|c|c|c|c|c|c|c|c}
            \hline
            & \multicolumn{3}{c|}{Component} & \multicolumn{6}{c}{Metric}\\
            \hline
            Model& EMSA& MAA& AAA& AP (\%) & mTTA (s) & TTA@R80  & FPS (s)& FLOPs (Ms) & Params (M)\\
            \hline
            A & \( \times \) & \( \bullet \) & \( \bullet \) & 85.60 & \underline{4.44} & 3.98  & 665.79& 298.39&3.05
\\
            B & \( \bullet \) & \( \times \) & \( \bullet \) & 84.78 & 4.40 & \textbf{4.65}  & 1484.04& 607.73&\textbf{3.12}
\\
            C & \( \bullet \) & \( \bullet \) & \( \times \) & \underline{87.33} & 3.38 & 3.68  & \textbf{1671.90}& 576.43&\underline{3.06}
\\
            \textbf{LATTE} & \( \bullet \) & \( \bullet \) & \( \bullet \) & \textbf{89.74} & \textbf{4.49} & \underline{4.04}  & \underline{1508.47}& \textbf{608.34}&\textbf{3.12}\\
            \hline
        \end{tabular}
    \end{table}

\subsection{Visualization}
To demonstrate LATTE's predictive capabilities, we analyze visualizations of its outputs across both accident-positive and accident-negative scenarios, supported by temporal probability analyses. Through comparative case studies, these visualizations elucidate the system's capacity to identify critical events and generate timely warnings, while also revealing operational constraints under specific edge-case conditions.

\begin{figure}
    \centering
    \includegraphics[width=1\linewidth]{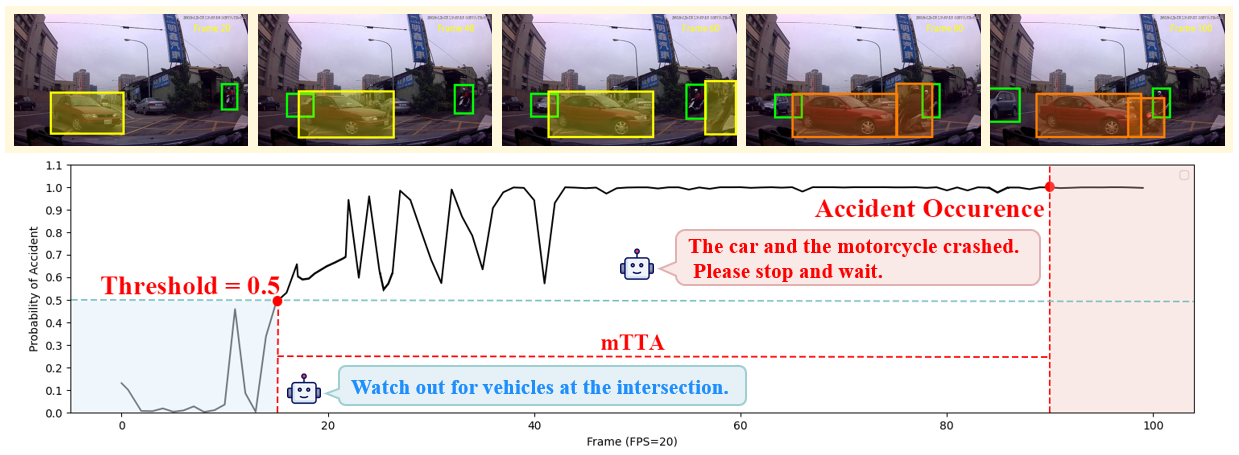}
    \caption{Anticipation of an accident-positive scenario. LATTE predicts the accident 3.7 seconds prior to its occurrence, with green bounding boxes denote the unrelated-accident objects, yellow bounding boxes mark the accident-related objects and orange bounding boxes highlight the accident participants at the actual moment of accident occurrence. The probability plot shows the prediction surpassing the 0.5 threshold, supported by FAA’s verbal alert.}
    \label{fig:fig2}
\end{figure}

Figure~\ref{fig:fig2} depicts an accident-positive scenario where a red car executing a left turn collides with an oncoming motorcycle at an intersection. LATTE accurately predicts the accident 3.7 seconds prior to impact, enabling critical intervention time. The model precisely identifies accident-critical objects (yellow bounding boxes) while filtering non-essential elements, such as the background motorcycle indicated in green. During frames 80-100, LATTE progressively highlights accident-prone objects in orange, with color intensity escalating as accident risk increases. Complementing this visual feedback, the FAA module triggers voice alerts when accident probability exceeds predefined thresholds, delivering timely warnings to vehicle occupants.

\begin{figure}
    \centering
    \includegraphics[width=1\linewidth]{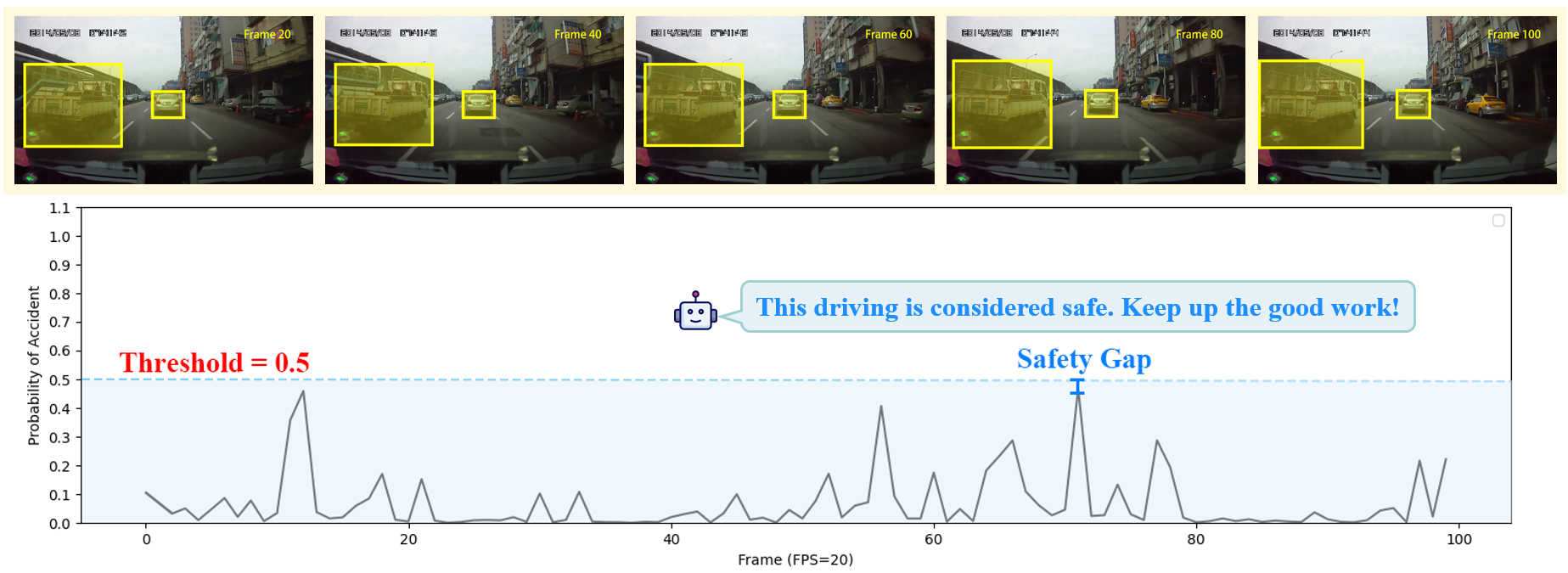}
    \caption{Anticipation of an accident-negative scenario. LATTE correctly maintains a low accident probability as the main vehicle navigates safely. Peaks in predictions around frames 20 and 70 are attributed to the proximity of a delivery truck but resolve as the risk decreases.}
    \label{fig:fig3}
\end{figure}

In contrast, Figure~\ref{fig:fig3} analyzes an accident-negative scenario captured under low-light evening conditions. The subject vehicle maintains center-lane positioning on a three-lane roadway, with positional relationships including a left-adjacent delivery truck and a leading white sedan. LATTE registers transient risk elevations at frames 20 and 70—attributable to the delivery truck's close proximity temporarily amplifying accident potential. This temporal pattern reverses as the truck diverges, with accident probability subsiding proportionally to inter-vehicle distance, demonstrating the system's responsive adaptation to evolving traffic dynamics. The visualization incorporates a ``Safety Gap'' metric, where sub-threshold probability values correspond to operationally safe states, thereby validating LATTE's equilibrium between spurious alert mitigation and rigorous safety evaluation.

\newcommand{\wideimgwidth}{0.95\linewidth}

\begin{figure}
    \centering
    \includegraphics[width=1\linewidth]{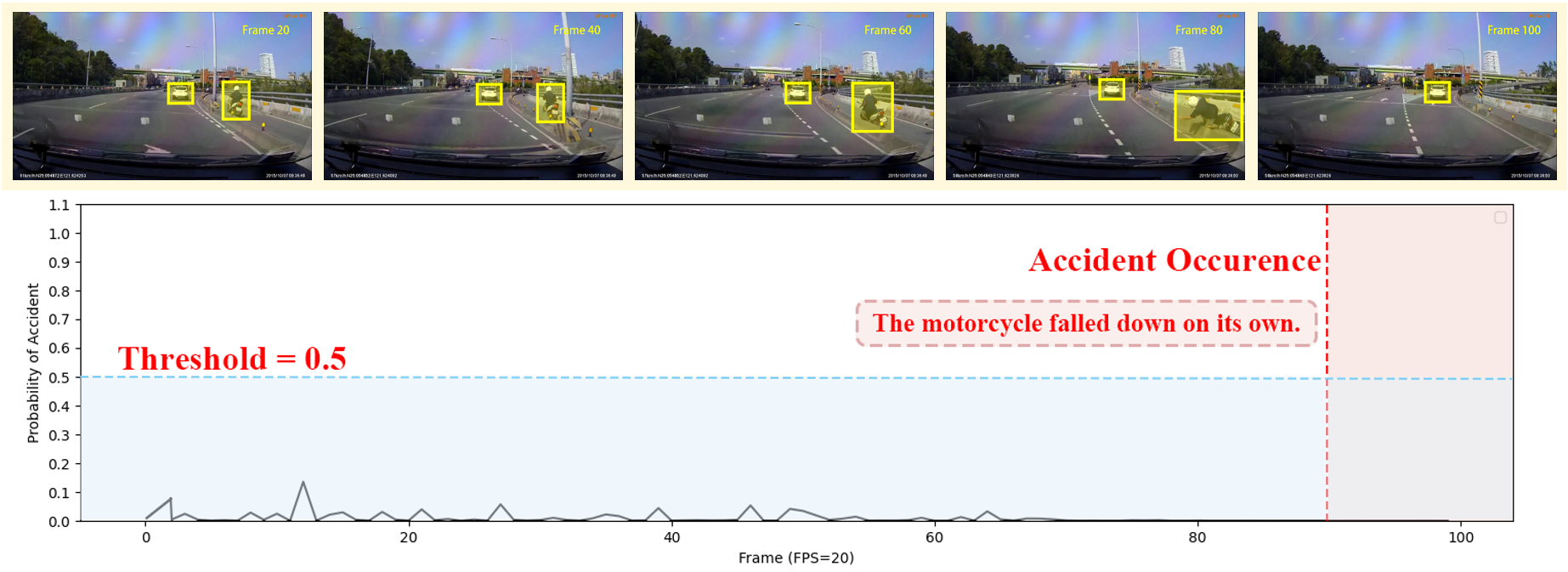}
    \caption{Failure case for accident anticipation. LATTE fails to predict a motorcycle crash due to limited training data for vehicle-infrastructure accident and a lack of surrounding traffic complexity. The probability plot remains below the 0.5 threshold.}
    \label{fig:fig4}
\end{figure}

To examine LATTE's limitations, Figure~\ref{fig:fig4} analyzes a failure case involving the model's erroneous low-probability anticipation for a motorcycle-barrier accident. The scenario features a motorcycle traversing a segregated non-motorized lane, exhibiting pronounced lateral oscillations from frame 60 until barrier impact at frame 90. Despite these kinematic precursors, LATTE's failure to anticipate the crash appears attributable to insufficient training data coverage for vehicle-infrastructure accident patterns. The lack of proximate vehicular interactions may have further compounded the error, given the model's inherent reliance on multi-agent dynamics as accident probability indicators.

These observations emphasize the critical need for expanding training dataset diversity to encompass under-represented vehicle-infrastructure conflict scenarios. Future LATTE iterations could benefit from enhanced multimodal sensing capabilities, particularly through the adoption of vehicular pose estimation to capture pre-collision kinematic patterns, while exploring driver state analysis via behavioral and physiological indicators (e.g., postural dynamics, gaze patterns, vigilance fluctuations) for early anticipation of human-factor risks such as fatigue-induced impairment or sudden medical anomalies. Such multimodal improvements would significantly strengthen LATTE's operational robustness in heterogeneous real-world environments.

\section{ Conclusion }
The LATTE framework demonstrates that efficient accident anticipation need not compromise accuracy, achieving this balance through four core components: EMSA for hierarchical spatial feature extraction, MAA for temporal dependency modeling, AAA for latent temporal dependencies, and FAA for human-intuitive alert generation. Comprehensive evaluations validate its dual strengths in early risk anticipation (demonstrated through AP/mTTA metrics) and operational efficiency (quantified via FLOPs/FPS measurements), establishing new state-of-the-art performance while maintaining minimal computational overhead. The FAA subsystem further enhances practical utility through semantically transparent feedback, bridging technical transparency and human-machine collaboration.

Persisting challenges emerge in highly dynamic urban ecosystems where multi-agent interactions, transient environmental conditions, and hardware-software degradation cycles strain real-time predictive fidelity. Notably, sustained operational risks including sensor calibration drift from hardware aging and module de-synchronization due to software updates may progressively destabilize inter-component collaboration, potentially compromising long-term accuracy. LATTE's current architecture exhibits sensitivity to photometric variations (e.g., low-light transitions, interference noise) and cross-modal dependencies between visual perception and linguistic feedback. Future research could explore hybrid sensor fusion architectures combining LiDAR, radar, and V2X data to enhance spatiotemporal awareness, potentially integrated with sparse attention mechanisms for improved computational efficiency. Edge-optimized model distillation methods might further address sustainable deployment constraints, while joint optimization of domain-adaptive visual processing and uncertainty-aware language generation could strengthen robustness against real-world operational variances. These advancements, however, may require concomitant development of prognostic health monitoring systems and version-controlled update protocols to mitigate lifecycle synchronization challenges between evolving hardware platforms and algorithmic frameworks. These advancements aim to reconcile computational efficiency with the stochastic complexity of real-world traffic environments, ultimately fostering resilient accident anticipation systems for autonomous vehicles.
        
\section{Acknowledgments}
This research is supported by Science and Technology Development Fund of Macau SAR (0021/2022/ITP), Shenzhen-Hong Kong-Macau Science and Technology Program Category C (SGDX20230821095159012), State Key Lab of Intelligent Transportation System (2024-B001), Jiangsu Provincial Science and Technology Program (BZ2024055), and University of Macau (SRG2023-00037-IOTSC, MYRG-GRG2024-00284-IOTSC).

\printcredits
\bibliographystyle{cas-model2-names}
\bibliography{main}
\end{document}